\documentclass[aps,pra,superscriptaddress,showpacs,twocolumn,10pt]{revtex4-1}
\usepackage[colorlinks=true,linkcolor=blue,urlcolor=blue,citecolor=blue]{hyperref}
\usepackage{amssymb}
\usepackage{afterpage}
\usepackage{graphicx}
\usepackage{amsmath}
\usepackage[space]{grffile}
\usepackage[T1]{fontenc}
\usepackage{braket}
\usepackage{color}
\usepackage[normalem]{ulem}

\newsavebox\CBox

\begin{document}

\title{Spin mixing in Cs ultralong-range Rydberg molecules: a case study}

\author{Samuel Markson}
\affiliation{%
 ITAMP,  Harvard-Smithsonian Center for Astrophysics 60 Garden St., Cambridge, MA 02138, USA}
 \affiliation{Physics Department, University of Connecticut, Storrs, CT 06269-3046, USA
}
 \email{smarkson@cfa.harvard.edu}

\author{Seth T. Rittenhouse}
\affiliation{%
 Department of Physics, the United States Naval Academy, Annapolis, Maryland 21402, USA}
\author{Richard Schmidt}
\affiliation{%
 ITAMP,  Harvard-Smithsonian Center for Astrophysics 60 Garden St., Cambridge, MA 02138, USA}
 \affiliation{Department of Physics, Harvard University, Cambridge MA 02138, USA}
\author{James P. Shaffer}
\affiliation{%
University of Oklahoma, Homer L. Dodge Department of Physics and Astronomy, Norman, Oklahoma 73072, USA}

\author{H. R. Sadeghpour}%

\affiliation{%
 ITAMP,  Harvard-Smithsonian Center for Astrophysics 60 Garden St., Cambridge, MA 02138, USA}

\date{\today}

\begin{abstract}
We calculate vibrational spectra of ultralong-range Cs$(32p)$ Rydberg molecules which form in an ultracold gas of Cs atoms. 
We account for the partial-wave scattering of the Rydberg electrons from the ground Cs perturber atoms by including the full set of spin-resolved $^{1,3}S_J$ and $^{1,3}P_J$ scattering phase shifts, and allow for the mixing of singlet ($S=0$) and triplet ($S=1$) spin states through Rydberg electron spin-orbit and ground electron hyperfine interactions. 
Excellent agreement with observed data in Sa{\ss}mannshausen {\it {\it et al.}} [Phys. Rev. Lett. {\bf 113}, 133201(2015)] in line positions and profiles is obtained. We also determine the spin-dependent permanent electric dipole moments for these molecules. This is the first such calculation of ultralong-range Rydberg molecules in which all of the relativistic contributions are accounted for.
\end{abstract}
\maketitle
\section{Introduction}
Rydberg atoms are weakly bound systems which can simultaneously exhibit quantum and classical behavior \cite{gallagher_rydberg_2005,buchleitner_non-dispersive_2002}. 
This quantum to classical evolution is at the heart of the Bohr-Sommerfeld quantization. Rydberg spectroscopy is a useful technique for probing many of the subtle properties of an atomic or molecular core, for measuring the Rydberg constant, and probing interactions in the surrounding gas or plasma. 
The interaction and collision of a Rydberg atom with the neutral and ionic species in the gas broadens and shifts the atomic lines, through which much can be learned about the scattering properties of the gas. 
The seminal measurements of Amaldi and Segr\`{e} \cite{amaldi_effetto_1934} confirmed that the classical macroscopic polarization of the dielectric medium was  insufficient to describe the Rydberg line shifts and that the scattering of Rydberg electrons from the perturber gas atoms would have to be accounted for. This led Fermi to develop a zero-range scattering theory, now called the Fermi pseudopotential (or contact potential) method \cite{fermi_sopra_1934}. 

Fermi realized that the low-energy scattering of a Rydberg electron from a perturber gas atom can be effectively described by a short-range elastic scattering interaction; this model has had much subsequent success in determining low-energy electron-atom scattering lengths of many other species \cite{thompson_pressure_1987}.
The form of zero-energy scattering of an s-wave electron from a perturber is then,
\begin{equation}
H_s(\vec{r},\vec{R})={2 \pi a_s(0)}\delta^{(3)}(\vec{r}-\vec{R})  
\label{eq:triplets}
\end{equation}
where $\vec{r}$ is the electronic coordinate, measured  from the Rydberg core, $\vec{R}$ is the vector connecting the Rydberg nucleus to the perturber atom nucleus, and $a_s(0)$ is the zero-energy s-wave scattering length. 

The delta-function contact interaction formalism can be extended to higher scattering angular momenta; an analytical form for p-wave scattering was derived by Omont  \cite{omont_theory_1977}, as 
\begin{equation}
H_{p}(\vec{r},\vec{R}) = 
6 \pi a_{p}^3(k)\delta^{(3)}(\vec{r} - \vec{R}) \overleftarrow{\nabla} \cdot \overrightarrow{\nabla} 
\label{eq:pwaveham}
\end{equation}
where $a_{p}(k)$ is the $k$-dependent p-wave scattering length

\begin{equation}
a_{\ell_{sc}}^{2 \ell_{sc} +1} (k) = -\frac{\tan{\delta_{\ell_{sc}}(k)}}{k^{2\ell_{sc} +1}}.
\label{eq:sphase}
\end{equation}
Here $\ell_{sc}$ is the orbital angular momentum about the perturber atom and $\ell_{sc} =0,1$ denote s- and p-wave scattering of the electron from the perturber atom, respectively.

The Born-Oppenheimer (BO) potential curves, as eigenstates of the Hamiltonian with ${H}_s$ and $H_p$ contributions, are highly oscillatory in internuclear distance $R$ due to the admixture of Rydberg electron wave functions with high principal quantum numbers $n$. 
It was realized in \cite{greene_creation_2000,hamilton_shape-resonance-induced_2002}  that those multi-well BO potentials can support bound vibrational levels when $a_{s}(0) <0$ as is the case for all alkali-metal atoms. 

\begin{figure}[t]
\includegraphics[width=\linewidth]{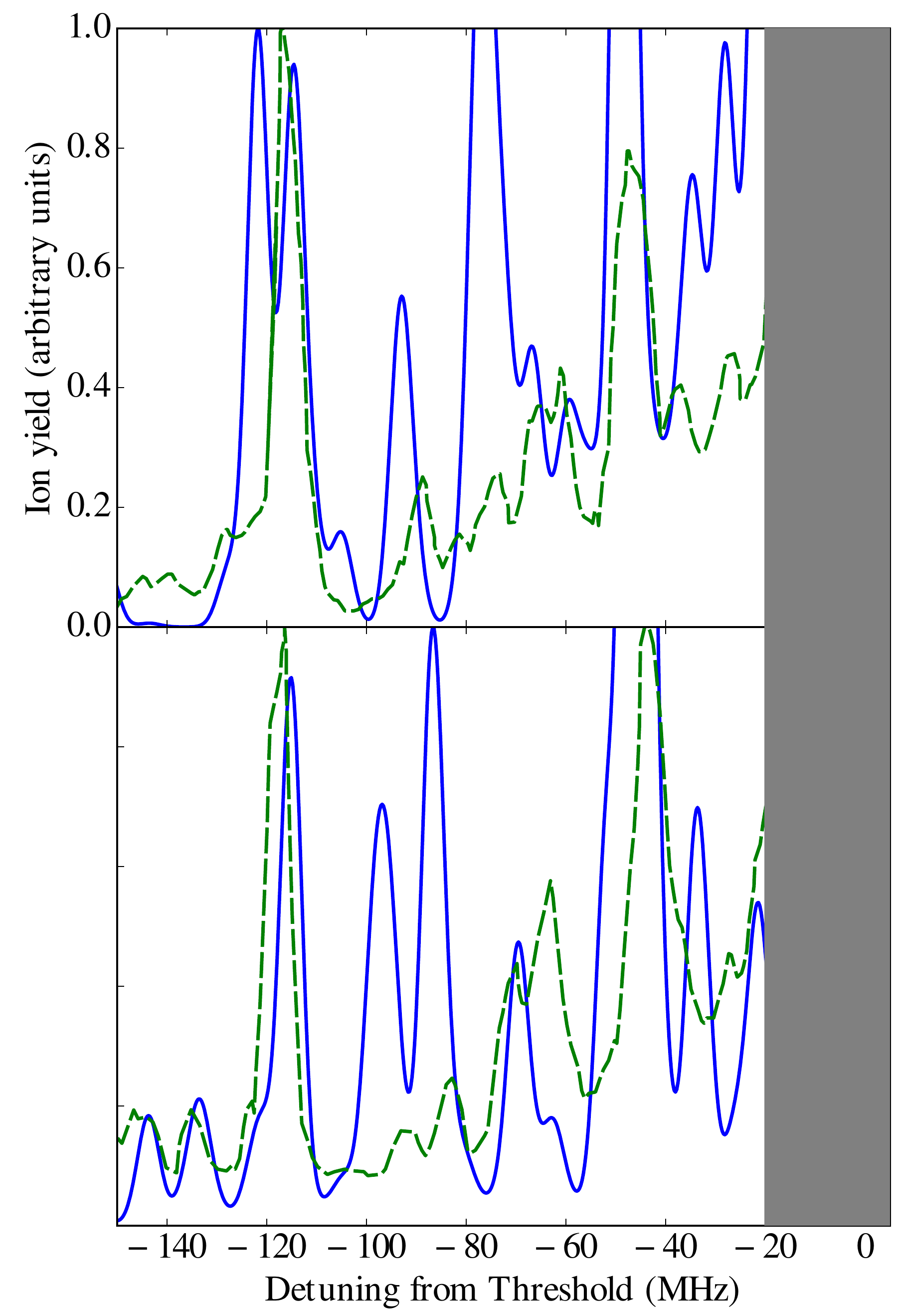}
\caption{The spectral line profiles are calculated (solid lines) and compared with observed spectral line features (dashed lines) in the F=4 (a) and F=3 (b) manifolds. See text for a description of the calculated line profiles. The threshold zero energy is the atomic Rydberg level Cs($32p_{3/2}$). The experimental data are from Ref.  \cite{sasmannshausen_experimental_2015}. }
\label{fig:fullspectra}
\end{figure}

Such exotic molecular Rydberg states were realized in magnetic and dipole traps, first in an ultracold gas of Rb atoms \cite{bendkowsky_observation_2009}, where such Rydberg molecules have $^3\Sigma\,\text{Rb}_2(ns)$ spherical symmetry.
It was confirmed that, even though such molecules were homonuclear, the mixing of Rb($ns$) levels with Rb$(n-3, l>2)$ hydrogenic manifolds, produces appreciable permanent electric dipole moments in these molecular species \cite{li_homonuclear_2011}. 
The prediction for Rydberg molecules with kilo-Debye dipole moments (trilobite molecules) were realized with ultracold Cs atoms \cite{tallant_observation_2012,booth_production_2015} in which the Cs$(n-4, l>2)$ degenerate manifolds are energetically much closer to Cs($ns)$ levels, hence providing for much stronger mixing of opposite parity electronic states. More recently, butterfly molecules (Rydberg molecules stemming from the presence of p-wave resonances) were predicted \cite{khuskivadze_adiabatic_2002,fabrikant2002,hamilton_shape-resonance-induced_2002} and confirmed \cite{niederprum_observation_2016}. 
In increasingly dense gases, additional molecular lines stemming from the formation of trimers, tetramers, pentamers, etc. have been observed \cite{gaj_molecular_2014,schlagmuller_ultracold_2016, schmidt_mesoscopic_2016}. 

The above scattering formalism is spin-independent, i.e. while the scattering phase shifts depend separately on the total spin of electrons, the scattering amplitudes add up incoherently. 
For Rydberg excitation in a gas of alkali-metal atoms, the total spin channels are $\text{S}=|\mathbf{s}_r+\mathbf{s}_g|=0, 1$ for singlet and triplet scattering, respectively, with $\mathbf{s}_r$ and $\mathbf{s}_g$ the Rydberg electron and the perturber ground electron spins. 

In this work we will account for all of the relevant and relativistic effects in Rydberg perturber atom scattering. We include all of the fine structure resolved s- and p-wave scattering Hamiltonians, the Rydberg electron spin-orbit, and the ground electron hyperfine interactions.  From the resulting BO potentials we not only predict the spatial structure and energies of the Rydberg molecules but can also reproduce the spectral line profiles in the recent experiment \cite{sasmannshausen_experimental_2015} as shown in Fig.~\ref{fig:fullspectra}.

\section{Hamiltonian}

The total Hamiltonian with all spin degrees of freedom included is,
\begin{multline}
H=H_0 + H_{{p}, T}\cdot P_T + H_{{p},S}\cdot P_S + \\
H_{{p}, T}\cdot P_T + H_{{p},S}\cdot P_S + H_{so} + H_{hf}
\label{eq:totalH}
\end{multline}
where $H_0$ is the Hamiltonian for the unperturbed Rydberg atom, and $H_{{(s,p)}, T}$ and $H_{{(s,p)}, S}$ are the (s-wave, p-wave) scattering Hamiltonians for triplet and singlet spin configurations, respectively. The operators $P_T=\mathbf{s}_r\cdot \mathbf{s}_g +3/4 $ and $P_S= 1 - P_T$ are the triplet and singlet projection operators for the total electronic spin.

It was first pointed out by Anderson {\it et al.} \cite{anderson_angular-momentum_2014} that the ground state hyperfine interaction can mix singlet and triplet spin configurations. The hyperfine Hamiltonian is ${H}_{hf}=A_{hf} \mathbf{s}_g \cdot \mathbf{i}_g$, where $\mathbf{i}_g$ is the nuclear spin and $A_{hf}$ is the hyperfine interaction; in Cs, the focus of this work, $A_{hf}$ = 2298.1579425 MHz, and $i_g=7/2$. Anderson {\it et al.} demonstrated this in Rb, where they observed bound vibrational levels due to mixing of singlet and triplet spins, even though the singlet zero energy scattering length for Rb is known to be small and positive. This is particularly important as, in magneto-optical and magnetic traps, spin alignment dictates that the interactions occur via the triplet scattering channel. 

The spin-orbit interaction for the Rydberg electron, $H_{so} = A_{so} \mathbf{l}_r\cdot\mathbf{s}_r$, where $\mathbf{l}_r$ is the orbital momentum and $A_{so}$ is the spin-orbit strength for $l_r\neq 0$ levels. 
Anderson {\it et al.} neglected this term, because it scales as $n^{-3}$.
For intermediate $n$, however, the spin-orbit splitting will be comparable to the hyperfine splitting.
The details of the matrix elements of $H_{so}$ and the terms for Cs(np) states will follow below. 

Recent studies have explored the extent to which spin effects are necessary to properly predict, {\it ab initio}, the vibrational spectra of these molecules, largely concluding that fine, hyperfine, and p-wave effects can have significant effects on these spectra \cite{anderson_angular-momentum_2014,sasmannshausen_experimental_2015}.
To date, no study has incorporated all of the interaction terms (s-wave, p-wave, spin-orbit, and hyperfine) on the vibrational spectra of Cs Rydberg molecules. Due to the large hyperfine shift in ${}^{133}$Cs, and the existence of several p-wave resonances at intermediate energies, these contributions can be significant in Cs.  
The current results are employed here to interpret the observations by Sa{\ss}mannshausen {\it et al.} \cite{sasmannshausen_experimental_2015}.  

\section{Hamiltonian Matrix Elements}\label{sec:MatE}
The matrix elements of the unperturbed Rydberg Hamiltonian ($H_0$) are calculated from the solutions $\phi_{nl_rm_r}(\vec{r})$ to the equation $H_0\phi_{nl_rm_r}(\vec{r}) = -\frac{1}{2 (n - \mu_{l_r})^2} \phi_{nl_rm_r}(\vec{r})$, where the quantum defects $\mu_{l_r}$ for Cs atom levels are used as follows: 
\begin{center}
\begin{tabular}{| c | c |}
\hline
$l_r$ & $\mu_{l_r}$ \\ \hline
0 & 4.05739 \\ \hline
1 & 3.57564 \\ \hline
2 & 2.471396 \\ \hline
3 & 0.0334998 \\ \hline
4 & 0.00705658 \\ \hline
$\geq$ 5 & 0  \\ \hline
\end{tabular}
\end{center}

\begin{figure}
\includegraphics[width=0.45\textwidth]{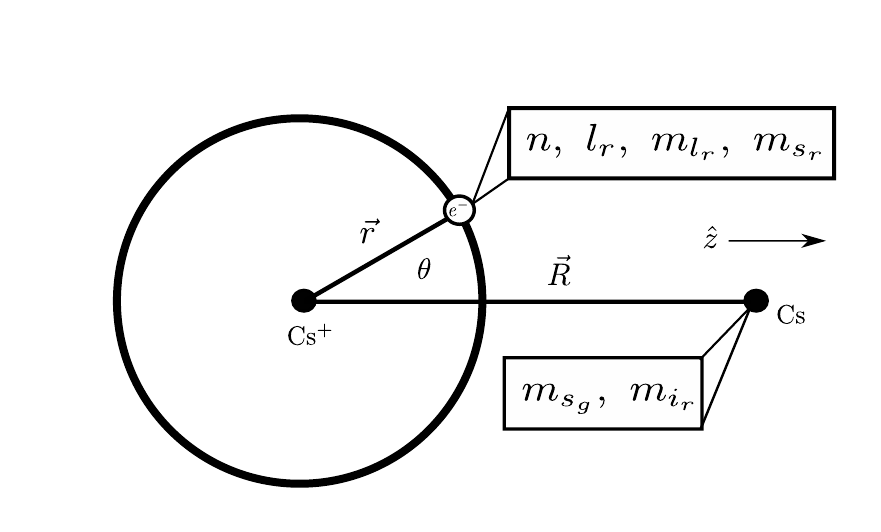}
\caption{The coordinate system used throughout the text.  Boxes show the six quantum numbers describing the basis set.  The electron spins for the Rydberg and ground state atoms are: $s_r$, $s_g$, while the  ground state atom nuclear spin $i_g$. Only the magnitude of the total angular momentum $\mathbf{K} = \mathbf{l}_r + \mathbf{s}_r + \mathbf{s}_g + \mathbf{i}_r$ and its projections are in general a good quantum numbers. }
\label{fig:coordinatesystem}
\end{figure}

We use the spherical coordinate system centered at the Rydberg core, as portrayed in Fig. \ref{fig:coordinatesystem}.  
The Rydberg orbitals forming our truncated basis set are comprised of $\{ns,(n+1)s, np, (n-1)d, (n-3)l_r\geq 3, (n-4)l_r\geq 3\}$ Rydberg wave functions.

The asymptotic form for the radial wave function is given by
\begin{equation}
F_{n l_r}(r) \propto r^{l_r} e^{-\frac{r}{n-\mu_{l_r}}} {}_1F_1\left(l_r-n+1+\mu_{l_r},2+2l_r,\frac{2 r}{n-\mu_{l_r}}\right)
\label{eq:hypergeo}
\end{equation}
where ${}_1F_1$ is the confluent hypergeometric function. For $\mu_l=0$, this reduces to the usual hydrogenic solution:
\begin{equation}
F_{n l_r} (r) = \sqrt{\left(\frac{2}{n}\right)^3 \frac{(n - l_r -1)!}{2 n (n+l_r)!}} e^{-\frac{r}{n}} \left(\frac{2 r}{n}\right)^{l_r} L_{n - l_r -1}^{(2 l_r +1)}\left(\frac{2 r}{n}\right)
\label{eq:laguerre}
\end{equation}
where $L_n^{(\alpha)}$ is the generalized Laguerre polynomial.  For non-integer quantum defects, the  wave functions $F_{nl_r=1,2}$ diverge at origin. To remedy this problem, they are matched  to  numerically calculated wave functions at small-$r$ \cite{marinescu_dispersion_1994}. 

\paragraph*{\textbf{Spin-dependent s-wave interaction.}---}
The Hamiltonian matrix elements describing s-wave scattering between basis functions $\ket{\phi_i}$ and $\bra{\phi_j}$ are,
\begin{multline}
\braket{\phi_i |{H}_{s}| \phi_j }  = 2 \pi \sum_{S}  a_s^S(k)\bra{ \phi _{i}} \delta^{(3)} (\vec{R}-\vec{r}) \ket{\phi _{j}} \\
\braket{s_{r_i},m_{s_{r_i}},s_{g_i},m_{s_{g_i}}|S,m_{S_i}}\times \\
\braket{S,m_{S_j} | s_{r_j},m_{s_{r_j}},s_{g_j},m_{s_{g_j}}}
\label{eq:swave}
\end{multline}
where $\ket{\phi_i}$ is shorthand for the basis element $\ket{n_i l_{r_i} m_{l_{r_i}} m_{s_{r_i}} m_{s_{g_i}} i_{s_{g_i}}}$, and $m_{S_i}=m_{s_{r_i}}+m_{s_{g_i}}$ is the total spin projection along the internuclear axis.
The s-wave scattering length $a_s^S(k)$ has been generalized to accommodate the ($S=0,1$) singlet and triplet scattering lengths. 

\paragraph*{\textbf{ Spin-dependent p-wave interaction.}---}
Additional caution is necessary when dealing with the p-wave electron-atom scattering, which depends on the total electronic spin ${\mathbf S}={\mathbf{s}_r+\mathbf{s}_g}$ and angular momentum ${\mathbf J}={\boldsymbol{\ell}}_{sc} +{\mathbf S}$ centered on the perturber atom.  
The triplet ($S=1$) p-wave scattering phase shift in Cs exhibits a relatively large splitting for $J=$0,1,2. 
This is in contrast to previous studies in Rb where the triplet p-wave scattering length is treated as a single resonance. 
The resulting Cs p-wave scattering interaction operator thus takes the form
\begin{multline}
\hat{H}_{p}= 6 \pi \sum_{J,m_{J}; S,m_S}  (a_{p}^{J,S}(k) )^{3}\delta^{(3)} ( \vec{R}-\vec{r}) \overleftarrow{\nabla }\cdot \overrightarrow{\nabla } \\
\ket{Jm_{J};S m_S} \bra{Jm_{J}; m_S }
\label{eq:pwaveinteraction}
\end{multline}
where $S=0,1$ and $m_{J}$ is the projection of {\bf J} along the internuclear axis. 
In the uncoupled basis $\left\vert nl_rm_{l_r};Sm_{S}\right\rangle$, where $\left\vert
nl_rm_r\right\rangle $ is a Rydberg orbital and $\left\vert Sm_{S}\right\rangle $
is the total spin state of the two electrons, matrix elements of this
interaction take on the form
\begin{multline}\label{eq:p_matrix_el}
\bra{\phi_i} \hat{H}_p \ket{\phi_j} = (\bar{a}_{p}^{m_{\ell_{sc}} m_S})^3 (k) \delta_{m_{l_{r_i}} m_{l_{r_j}}} \delta_{m_{S_i} m_{S_j}}  \\
\cdot \bra{\psi_i } \delta^{(3)} ( \vec{R}-\vec{r}) \overleftarrow{\nabla }\cdot \overrightarrow{\nabla } \ket{\psi_j} 
\end{multline}
with the effective scattering volume as
\begin{multline}
(\bar{a}_{p}^{m_{\ell_{sc}} m_S})^3 (k) = \\
\sum_{J=0,1,2}\left( \braket{(\ell_{sc}=1) m_{\ell_{sc}} ,(S=1) m_S | J m_J}^2 (a_p^{J,S=1})^3 (k)\right)\cdot P_T + \\
\left(\braket{(\ell_{sc}=1) m_{\ell_{sc}} ,(S=0) m_S | (J=1) m_J}^2 (a_p^{J=1,S=0})^3 (k)\right)\cdot P_S
\label{eq:effscattvol} 
\end{multline}
where $\braket{ Lm_{\ell_{sc}},Sm_{S}|Jm_{J}} $ is a Clebsch-Gordan coefficient coupling the orbital angular momentum of the Rydberg electron to the combined total spin of the ground state atom and Rydberg electron. Note that
$m_{\ell_{sc}}=m_{l_r}$, since angular momentum projections are invariant under translationalong the axis of projection.
Because of the projection onto the p-wave relative angular momentum
states, only Rydberg states with spatial angular momentum projection $m_{l}=0
$ or $\pm 1$ contribute to the interaction. 

The spatial integral in Eq. \ref{eq:p_matrix_el} can be evaluated  as%
\begin{widetext}
\begin{multline}
\bra{ \phi _{i}} \delta^{(3)} ( \vec{R}-\vec{r}) \overleftarrow{\nabla }\cdot \overrightarrow{\nabla }\ket{ \phi _{j}}=\lim_{\vec{r}\rightarrow R\hat{z}}\Bigg\{ \dfrac{2l_{i}+1}{4\pi }\delta _{m_{l_{i}}0}\dfrac{\partial F_{n_{i}l_{i}}^{\ast }(r=R) }{\partial r}\dfrac{\partial F_{n_{j}l_{j}}(r=R) }{\partial r} \\
+\delta _{\vert m_{l_{i}}\vert 1}F_{n_{i}l_{i}}^{\ast}(r=R) F_{n_{j}l_{j}}(r=R) \int  d(\cos\theta) d\phi \nabla Y_{l_{i}m_{l_{i}}}^{\ast }( \theta ,\phi ) \cdot \nabla Y_{l_{j}m_{l_{j}}}( \theta ,\phi )  \Bigg\} \delta_{m_{l_{i}}m_{l_{j}}},
\end{multline}%
\end{widetext}
where $Y_{lm}( \theta ,\phi ) $ is a spherical harmonic, $%
F_{nl}(r) $ is the radial part of the Rydberg wave function, and we suppress the subindex $r$ on $(l,m)$ for notational convenience.
It should be noted that the radial derivatives only act on wave functions
with $m_{l}=0$ while the angular derivatives $\nabla Y_{l_{j}m_{l_{j}}}$ are only non-zero for states with $m_{l}=\pm 1$.

\paragraph*{\textbf{ S- and P-wave scattering phase shifts.}---} In order to fully characterize the electron-perturber interaction, the s- and p-wave scattering lengths must be determined.  In the case of $a^{J=0,1,2,S=1}_p$ and $a^{S=1,0}_s$, these are derived by solving the scattering equation including the polarization potential $V_{e^{-}\text{-}Cs} = -\alpha/2r^4$ where $\alpha=402.2$ a$_0^3$ is the polarizability of the ground state Cs atom. 
We extract the phase shift by enforcing a hard wall boundary condition at short range ($r_0\lesssim 3$ a.u.) on the polarization potential.  

\begin{figure}
  \includegraphics[width=0.45\textwidth]{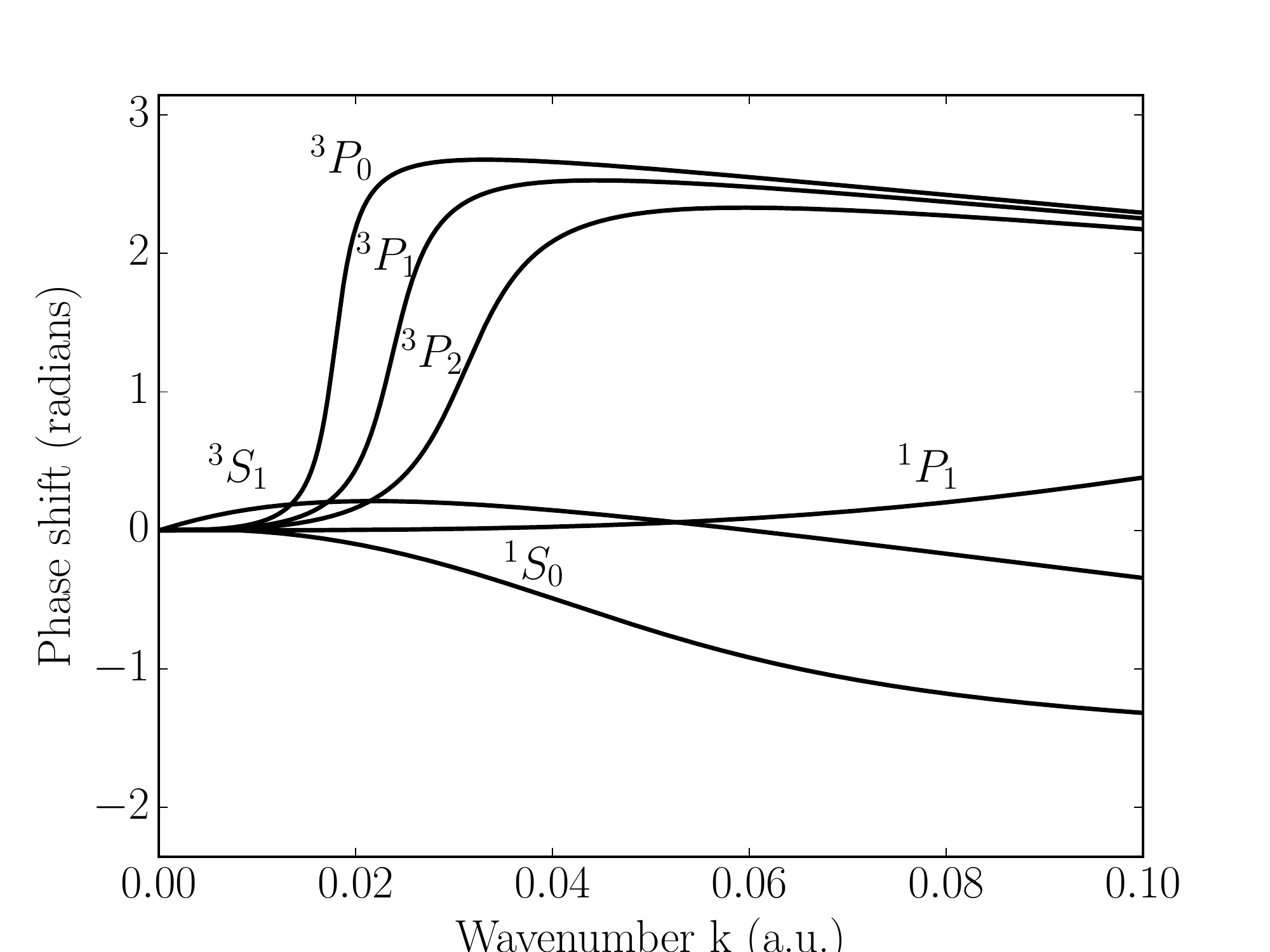}
\caption{The $^{1,3}S_J$, $^{1,3}P_J$ electron- Cs($6s$) phase shifts as a function of the electron scattering momentum are reproduced here. The $^1P_1$ phase shift was obtained from a calculation by U. Thumm \cite{uwethummcomm}. The $^{1,3}S_0$ and $^3P_{J=0,1,2}$ resonant phase shifts are obtained by matching a short distance oundary condition to experimental measurements of scattering resonances \cite{scheer_experimental_1998,marinescu_dispersion_1994}.
}
\label{fig:phaseshifts}
\end{figure}

For $a^{S=1,0)}_s$, we adjust the hard wall to reproduce  experimentally known zero-energy scattering lengths. For $a^{J=0,1,2,S=1}_p$, the position of the hard wall by is chosen so as to enforce a resonance in the scattering phase shift, i.e. $\delta_{E} = \pi/2$, at a resonant energy $E$ consistent with experimental measurements.  
While it is unlikely that this simple procedure captures all of the details of the p-wave electron-atom scattering process, the effects of the p-wave interaction on the molecular potentials are largely captured by the position and width of the $^3P_J$ scattering resonances.  
Specifically, we set the position of the $^3P_1$ to 8 meV, as observed \cite{scheer_experimental_1998}.  The $^3P_0$ and $^3P_2$ resonance positions are set to respectively be 3.8 meV below and 7.2 meV above the $^3P_1$ resonance position in accordance with \cite{thumm_evidence_1991}.  The energy-dependent phase shifts for  S=0, and 1 spin scattering in s-wave and p-wave are shown in Fig~\ref{fig:phaseshifts}.

\paragraph*{\textbf{Rydberg spin-orbit matrix elements.}---} The Rydberg electron spin-orbit Hamiltonian has the form $ {H}_{so}=A_{so}(n,l_r) \mathbf{l}_r \cdot \mathbf{s}_r$, whose coefficients for Cs(np) states have been measured in {\cite{goy_millimeter-wave_1982}}. The matrix elements are
\begin{equation}
\braket{\phi_i |{H}_{so}| \phi_j }=A_{so}(n,l_r) \mathbf{l}_r \cdot \mathbf{s}_r
\label{eq:spinorbit}
\end{equation}
where, for $l_r=1,2,3$ \cite{goy_millimeter-wave_1982}
\begin{align}
A_{so}(n,l_r)&=\nonumber\\
\left(l_r+\frac{1}{2}\right)^{-1} \big[&A(l_r)(n-\epsilon(n,l_r))^{-3} \nonumber\\
&+B(l_r) (n-\epsilon(n,l_r))^{-5} \nonumber\\
&+C(l_r) B(l_r) (n-\epsilon(n,l_r))^{-7}\big] 
\label{ml:spinorbitprefactor}
\end{align}
where $\epsilon(n,l_r)=\epsilon^{*}(l_r) + a(l_r) (n-\epsilon^{\*})^{-2}$

The $l_r$-dependent parameters $A,B,C, \epsilon^{\*}, a$ are
\begin{center}
\begin{tabular}{| c | c | c | c |}
\hline
 & $l_r=1$ & $l_r=2$ & $l_r=3$ \\ \hline
$A(l_r) \text{ (MHz)}$ & 2.13925e8 & 6.02183e7 & -9.796e5 \\ \hline
$B(l_r) \text{ (MHz)}$ & -5.6e7 & -5.8e7 & 1.222e7 \\ \hline
$C(l_r) \text{ (MHz)}$ & 3.9e8 & 0.0 & -3.376e7 \\ \hline
$\epsilon^{*} $ & 3.57531 & 2.47079 & 0.03346 \\ \hline
$a$ & 0.3727 & 0.0612 & -0.191 \\ \hline

\end{tabular}
\end{center}
while for $l_r \geq 4$, $A_{so}(n,l_r)$ takes on its hydrogenic values:
\begin{equation}
A_{so}(n,l_r)=\frac{\mu_0}{4 \pi} g_{l} \mu_{B}^2 \frac{1}{n^3 l_r (l_r + \frac{1}{2}) (l_r +1)} 
\label{eq:hydrogenicso}
\end{equation}
where $g_l$ is the Land\'{e} g-factor, $\mu_0$ is the vacuum permeability, and $\mu_B$ is the Bohr magneton.  

In the uncoupled angular momentum basis, the operator $\mathbf{l}_r \cdot \mathbf{s}_r$ has the representation~\cite{sakurai_modern_2011}
\begin{equation}
\mathbf{l}_r \cdot \mathbf{s}_r={l}_{z_r} {s}_{z_r} + \frac{1}{2} {l}_{{+}_r} {s}_{{-}_r} + \frac{1}{2} {l}_{{-}_r} {s}_{{+}_r}
\label{eq:ladders}
\end{equation}
where $({l},{s})_{r,\pm}$ are the ladder operators for the Rydberg electron orbital angular momentum and spin.  
This representation allows to determine the matrix elements between different combinations of angular harmonic $Y_{l_i,m_{l_i}}$ and spinors $\chi_{s_{i},m_{s_i}}$
\begin{equation}
\bra{Y_{l_i,m_{l_i}} \chi_{s_{i},m_{s_i}} } \mathbf{l}_r \cdot \mathbf{s}_r \ket{Y_{l_j,m_{l_j}} \chi_{s_{j},m_{s_j} } }.
\label{eq:matel}
\end{equation}
They are
\begin{multline}
\braket{\mathbf{l}_r \cdot \mathbf{s}_r} = \\
\left\{
	\begin{array}{ll}
		m_{l_i} m_{s_i}  & \mbox{if } l_i=l_j,s_i=s_j,\\
 &\ \ \ m_{l_i}=m_{l_j},m_{s_i}=m_{s_j} \\
		\frac{1}{2} \sqrt{l_i(l_i+1)-l_{z,i}(l_{z,i}\pm 1)}  \\ 
                 \cdot\sqrt{s_i(s_i+1)-s_{z,i}(s_{z,i}\mp 1)} & \mbox{if } l_i=l_j,s_i=s_j, \\
 &\ \ \ m_{l_i}=m_{l_j}-1,\\
 &\ \ \ m_{s_i}=m_{s_j}+1 \\
		\frac{1}{2} \sqrt{l_i(l_i+1)-l_{z,i}(l_{z,i}\mp 1)}  \\ 
               \cdot  \sqrt{s_i(s_i+1)-s_{z,i}(s_{z,i}\pm 1)} & \mbox{if } l_i=l_j,s_i=s_j, \\
 &\ \ \ m_{l_i}=m_{l_j}+1,\\
 &\ \ \ m_{s_i}=m_{s_j}-1
	\end{array}
\right.
\end{multline}
Through the ladder terms, the spin-orbit coupling will therefore couple $\Sigma$ and  $\Pi$- states, as well as singlet (S=0) and triplet (S=1) states.  

\paragraph*{\textbf{ Ground hyperfine matrix elements.}---} The ground electron hyperfine interaction Hamiltonian matrix elements are calculated as
\begin{equation}
\mathbf{s}_g \cdot \mathbf{i}_g=s_{z_g} i_{z_g} + \frac{1}{2} s_{+_g} i_{-_g} + \frac{1}{2} s_{-_g} i_{+_g}
\label{eq:laddershf}
\end{equation}
where $({s},{i})_{\pm,g}$ are the ladder operators for the perturber valence electronic and nuclear spins. 

We chose to demonstrate the utility of our method toward calculation of the vibrational spectrum of Cs$(6s_{1/2})$-Cs$(32p_{3/2})$ Rydberg molecules \cite{sasmannshausen_experimental_2015}. 
For this particular Rydberg excitation, the fine-structure splitting $\Delta E_{32p_{1/2}-32p_{3/2}}$ is nearly degenerate with the ground state hyperfine splitting, $\Delta E_{hf} \approx 9.2$ GHz. 

For $^{133}$Cs with ${i_g}=7/2$, therefore, our full basis, including angular momentum degrees of freedom, is 
\begin{align}
\{32s ,33s ,32p ,31d ,&n=29, 3 \leq l_r \leq 28,\nonumber\\
&n=28,3 \leq l_r \leq 27\}   \nonumber\\ 
&\times\left\{-l_r\leq m_{l_r}\leq l_r \right\}\nonumber \\
&\times \left\{ m_{s_r}=\pm \frac{1}{2}\right\} \nonumber \\
&\times\left\{ m_{s_g}=\pm \frac{1}{2}\right\} \nonumber\\
&\times \left\{ m_{i_g}=\pm \frac{1}{2}, \pm \frac{3}{2}, \pm \frac{5}{2}, \pm \frac{7}{2}\right\} .
\end{align}
We truncate this basis such that no basis element has $|m_{l_r}| > 2$
The total number of basis states included in our calculation is 8480.

The projection of the total angular momentum, ${\bf K}$ onto ${\bf R}$, $m_K=m_{l_r}+m_{s_r}+m_{s_g}+m_{i_g}$ is a good quantum number, so that the basis set diagonalization need only be performed in blocks of 1060, 1008, 851, 638, 422, 209, and 52 elements, for $|m_K|=\frac{1}{2},\frac{3}{2},\frac{5}{2},\frac{7}{2},\frac{9}{2},\frac{11}{2},\frac{13}{2}$, respectively.

\section{Results and Discussions}
\begin{figure}[t]
\includegraphics[width=\linewidth]{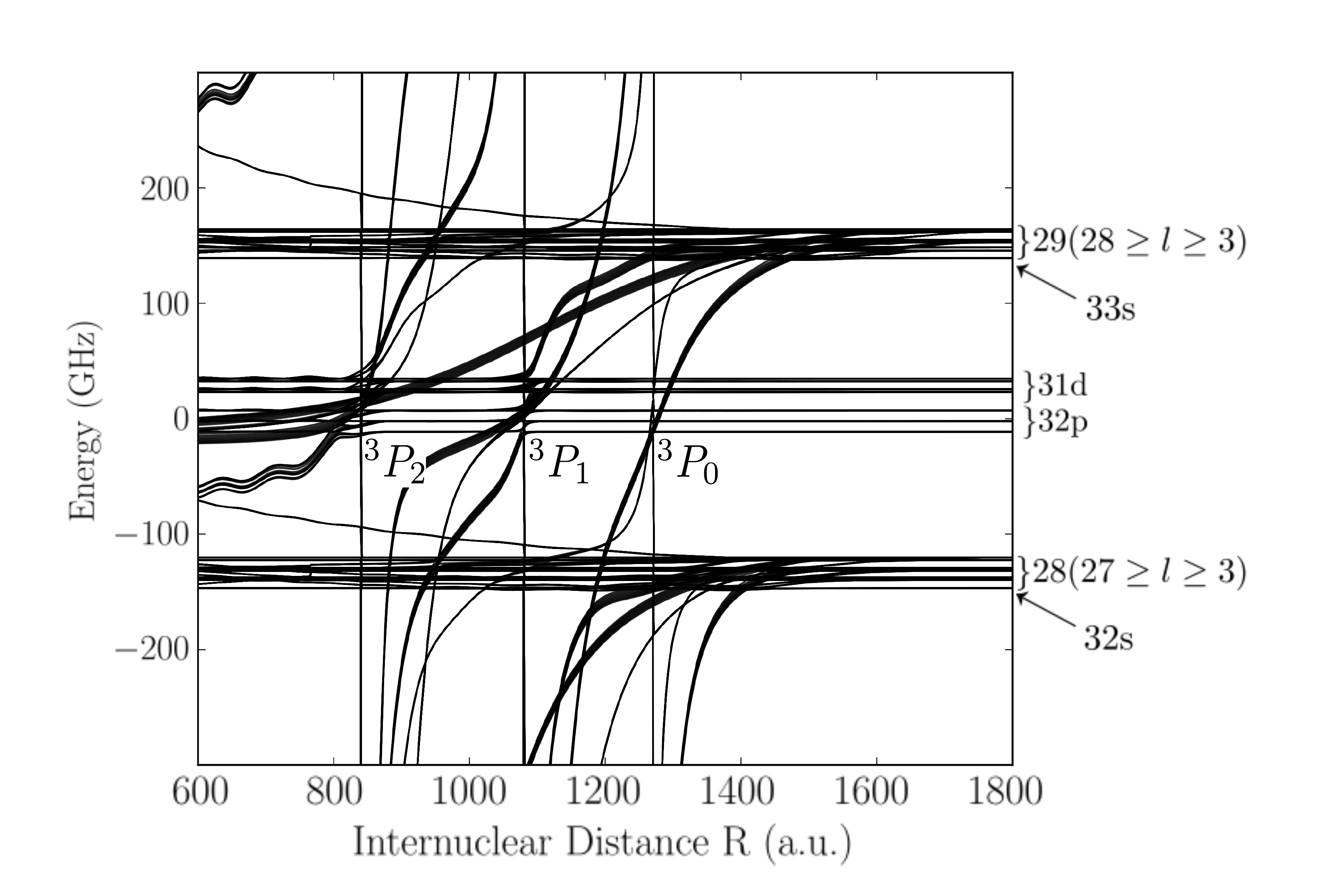}
\caption{Born-Oppenheimer potential energy curves.
The atomic Rydberg dissociation levels are indicated on the right side of the graph; two sets of degenerate hydrogenic manifolds are employed. Each Rydberg level is split due to SO and HF interactions in the Rydberg and ground states. On the scale shown, the Rydberg molecule energy landscape also highlights the dramatic influence of the $^3P_J$ resonances (labelled in the figure) which manifest themselves in the complicated set of avoided crossings. }
\label{fig:all}
\end{figure}

In Fig.~\ref{fig:all}, we show the full set of BO potential energy curves which result from the diagonalization of Eq.~\ref{eq:totalH} with the basis functions defined in Sec. \ref{sec:MatE}. This landscape of Rydberg potential energies reveals the  influence of the three $^3P_J$ resonances. Due to the different widths of J-resonances, see Fig.~\ref{fig:phaseshifts}, the avoided crossings of molecular potentials in $R$ occur at different locations with the various unperturbed Rydberg manifolds.  For example, near the 31d Rydberg level, the  $^3P_1$ resonance crosses before the narrower  $^3P_0$ resonance.

\begin{figure}[t]
\includegraphics[width=0.5\textwidth]{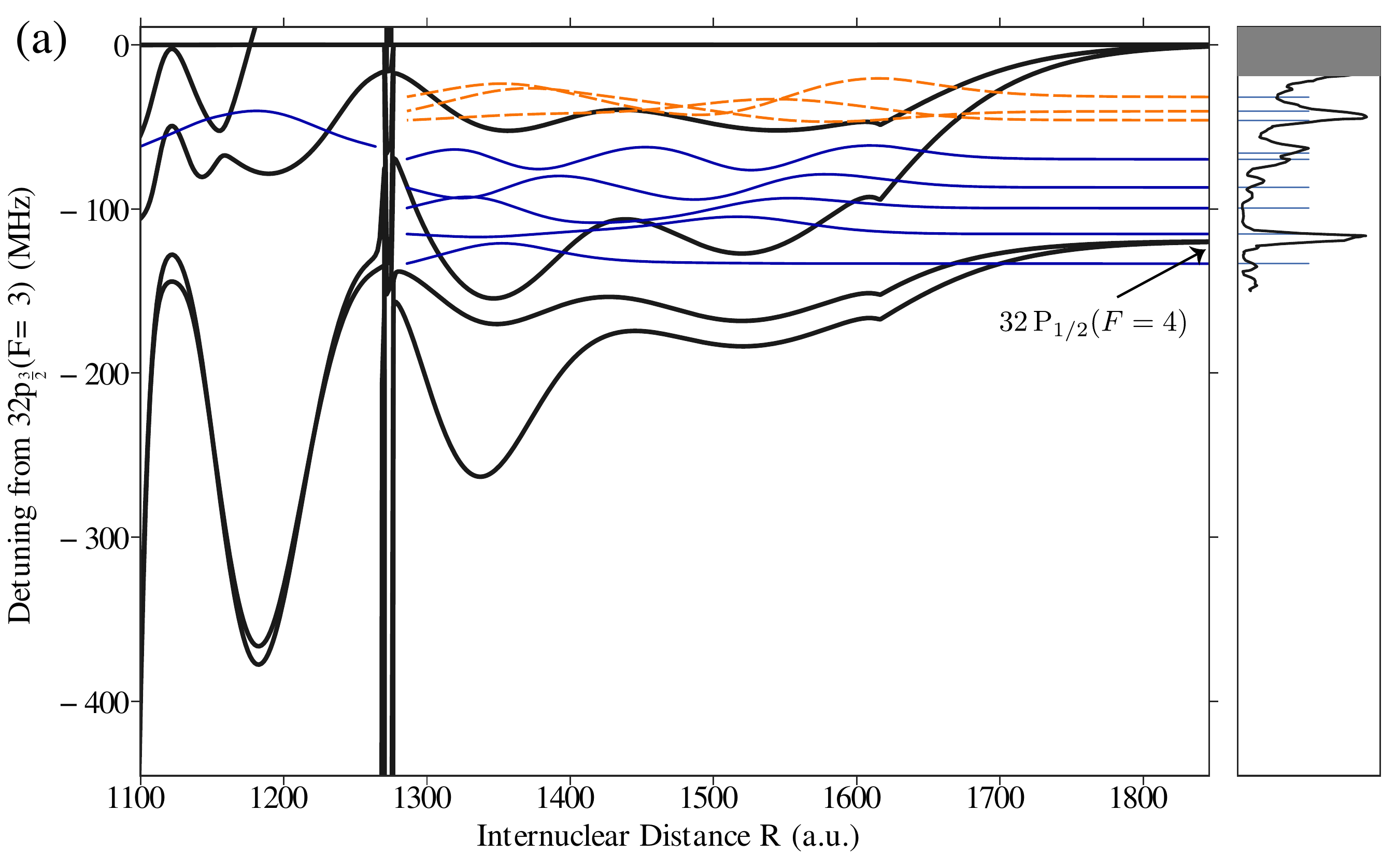}
\includegraphics[width=0.5\textwidth]{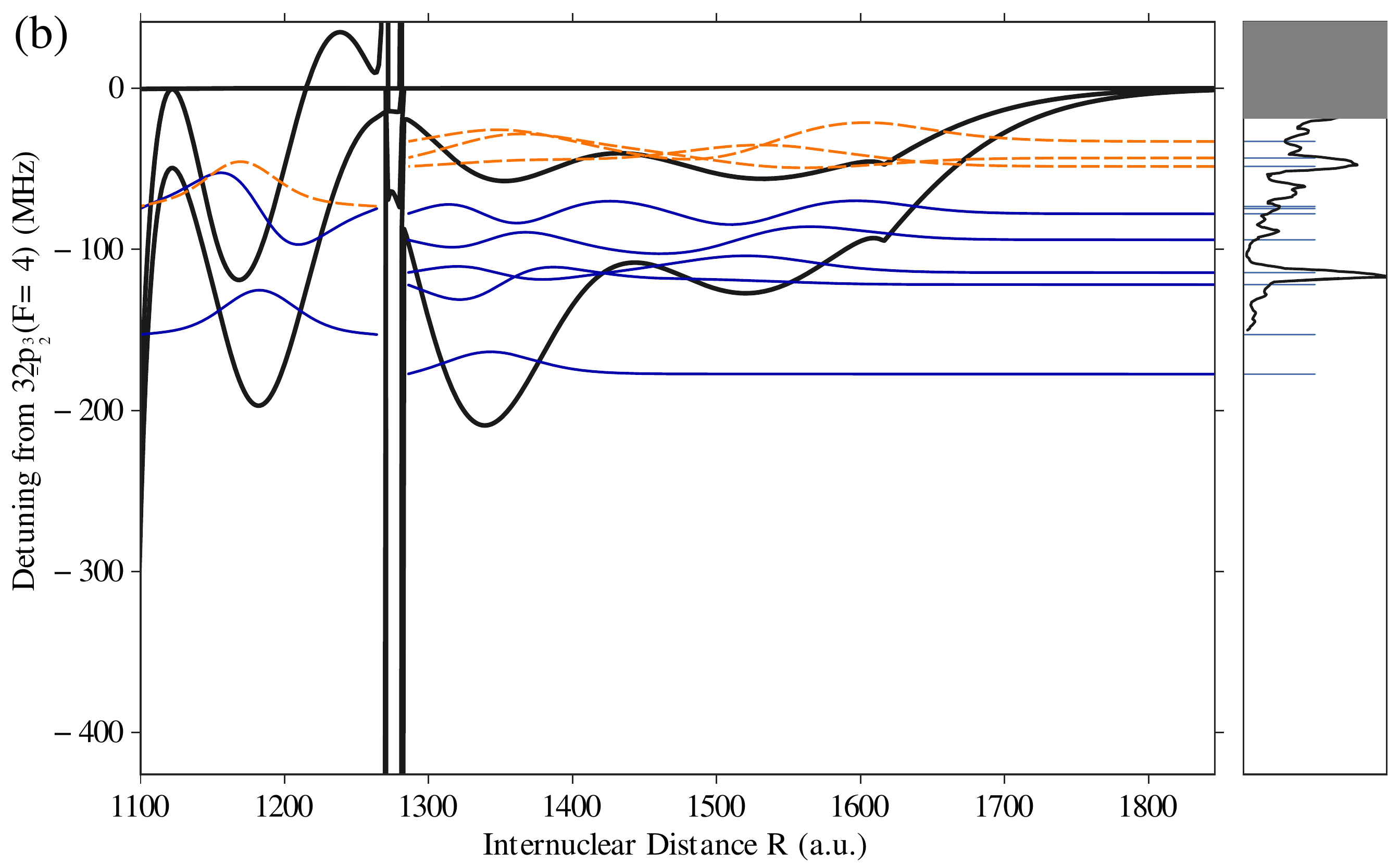}
\caption{(a) The BO potential energy curves correlating to the Cs$(6s$) - Cs($32p_{3/2}$) F=3 asymptote, for the total projection quantum number $m_K=\frac{1}{2}$. There are two sets of potential curves (solid black): the lower two curves correlate to the Cs($32p_{1/2}$) F=4 atomic threshold and are within $\sim 100$ MHz of the F=3 threshold because in Cs($32p$) excitation, the Rydberg spin-orbit  and the ground hyperfine splittings are nearly degenerate. The lowest curve in each set corresponds to the predominantly triplet symmetry and the upper curve in each set becomes sufficiently attractive to support vibrational levels (dashed lines) due to the mixing of triplet and singlet channels. The vertical lines are due to the presence of a narrow $^3P_0$ scattering resonance which crosses several atomic Rydberg levels; another crossing due to the $^3P_1$ scattering resonance is near $R\sim 1100 \ a_0$. In the F=3, $m_K=\frac{1}{2}$ manifold of states, the $\nu=0$ vibrational states in the predominantly triplet and mixed electronic potentials have, respectively, permanent electric dipole moments of 9.8 and 3.7 D; (b) the potential energy curves correlating to the  Cs$(6s$) - Cs($32p_{3/2}$) F=4 asymptote and the associated vibrational levels. The calculated vibrational energies  are shown as blue sticks on the right side panes, which illustrate the experimental absorption sprectra. 
}

\label{fig:mk12}
\end{figure}

While in Fig.~\ref{fig:all} the potential energy curves are shown for all possible projections $m_K$, in Fig.~\ref{fig:mk12} we show in detail the BO potentials for $m_K=1/2$. In the outer region, there are two distinct sets of curves; the lower set corresponds to the potential curves dissociating to the Cs$(6s_{1/2})$ - Cs$(32p_{1/2})$ $F=4$ threshold, and the upper set of curves dissociate to the Cs$(6s_{1/2})$ - Cs($32p_{3/2})$ $F=3$ threshold.  
The $F=3$ and $F=4$ thresholds  are within $\sim 100$ MHz of each other because in excitation of Cs($32p)$, the Rydberg SO and the ground HF splittings are nearly degenerate. 
Within each set, the lowest curve refers to a predominantly triplet potential energy curve and the upper curve refers to the more mixed singlet/triplet potential. 
Generally, in the outer region, defined by   internuclear distances $R$ greater than all the p-wave resonance crossings, the predominantly triplet curve will have greater than 90\% triplet character --- the singlet  mixing enters mainly through the $^1\Pi$  molecular symmetry --- while the mixed curve will have between 60\% and 70\% triplet character.
There will also be other non-binding potential energy curves  (up to two more in a given $m_K$ block) which largely have $\Pi$ character.
We stress that all of these molecular potentials are admixtures of $\Sigma$ and $\Pi$ symmetry states;
$\Delta$ contributions are in principle also present, but not of significance here. 
On the scale of  Fig.~\ref{fig:mk12} the relativistic $^3P_J$ scattering resonances manifest themselves as sharp vertical lines.

We calculate the bound vibrational wave functions in the Born-Oppenheimer approximation using the previously calculated potential energy curves. The resulting Rydberg molecular binding energies are indicated by the thin blue lines in the rightmost  pane of each figure, together with the absorption spectra measured in \cite{sasmannshausen_experimental_2015}. It is evident from the comparison that many of the spectral features in the experiment are reproduced in Fig.~\ref{fig:mk12}, which demonstrates that the experiment resolves different $m_K$ vibrational lines.

\begin{figure*}[p]
\centering  
\begin{minipage}{.5\textwidth}
  \centering
  \includegraphics[width=\linewidth]{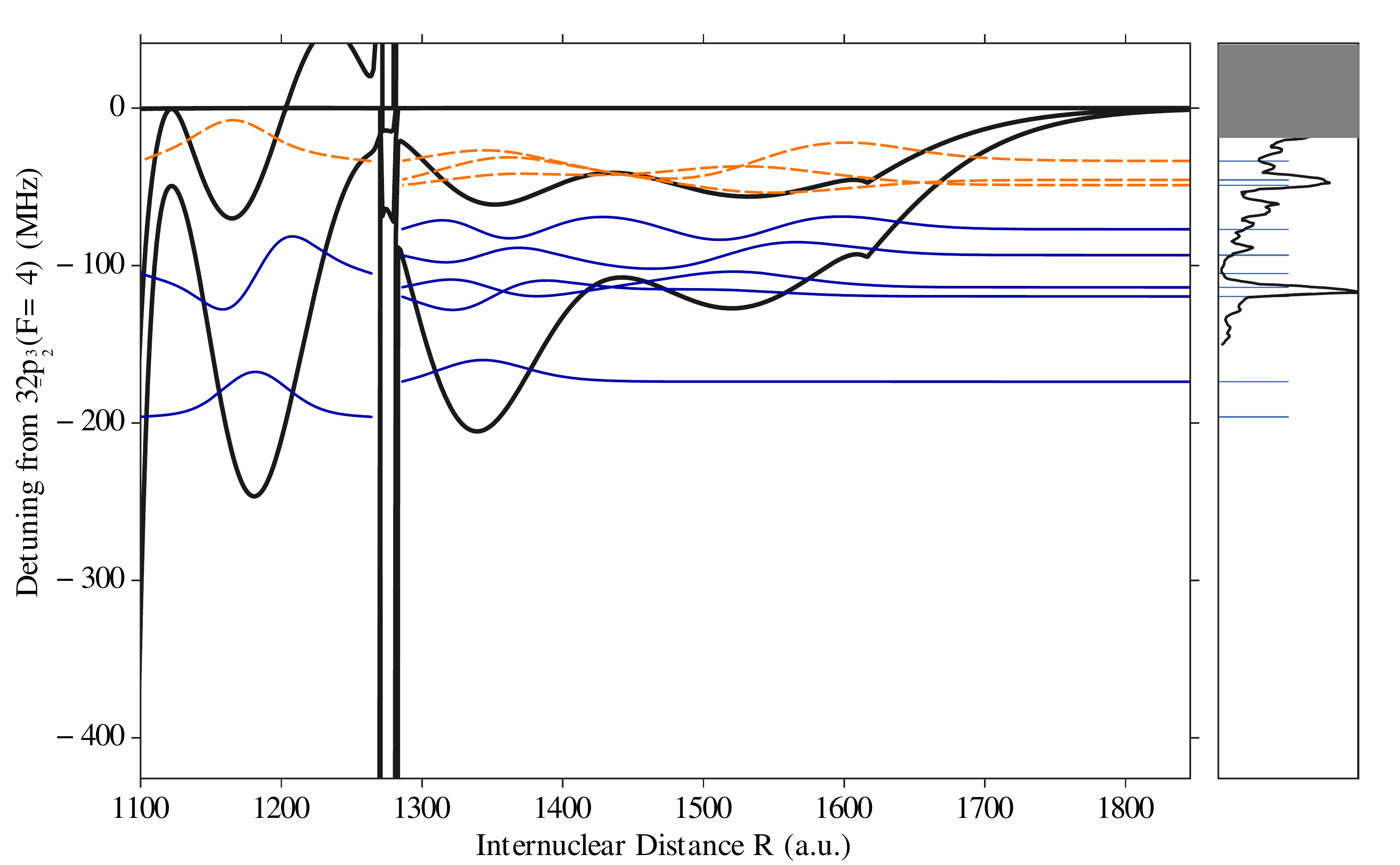}
\end{minipage}%
\begin{minipage}{.5\textwidth}
  \centering
  \includegraphics[width=\linewidth]{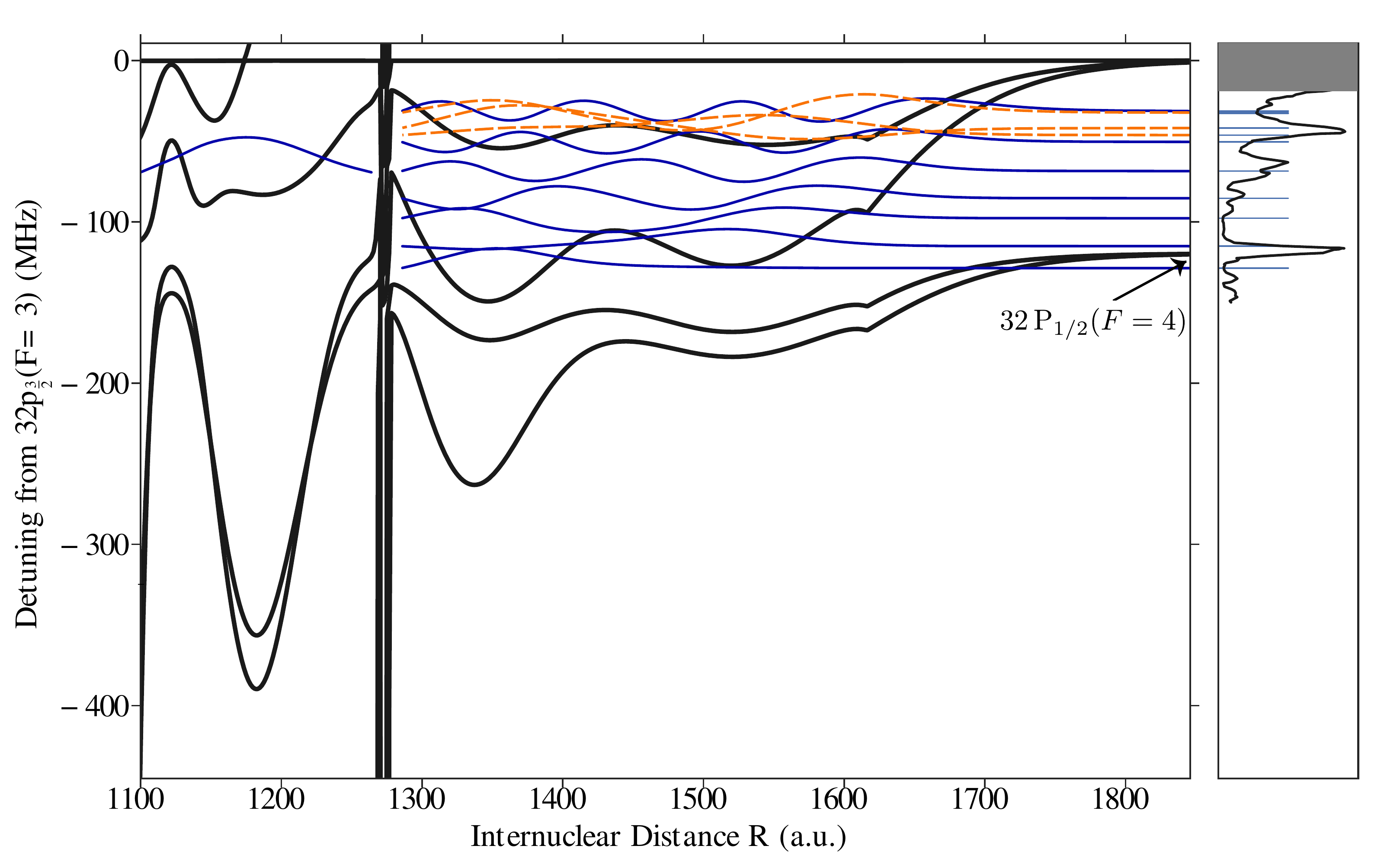}
\end{minipage}

\begin{minipage}{.5\textwidth}
  \centering
  \includegraphics[width=\linewidth]{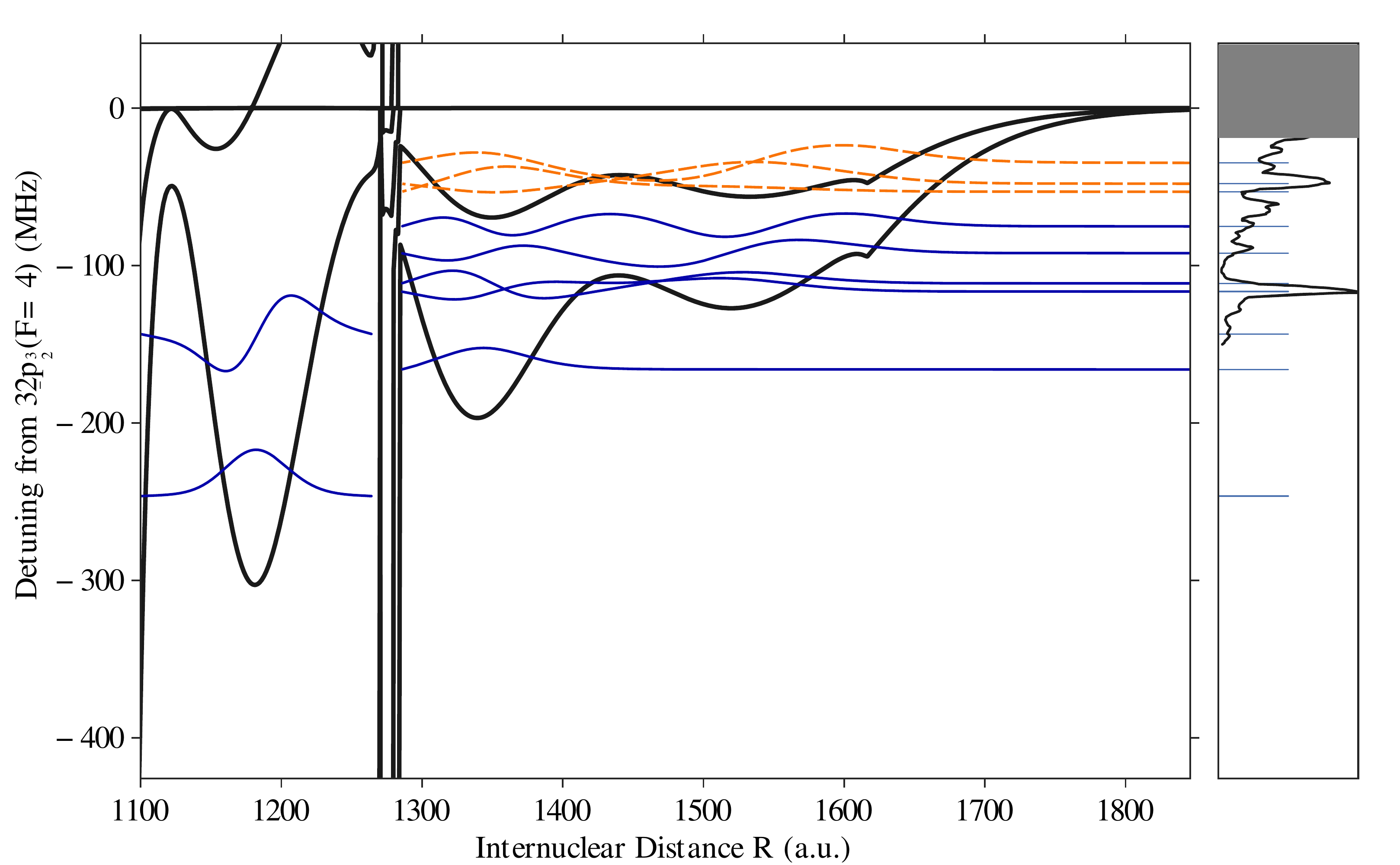}
\end{minipage}%
\begin{minipage}{.5\textwidth}
  \centering
  \includegraphics[width=\linewidth]{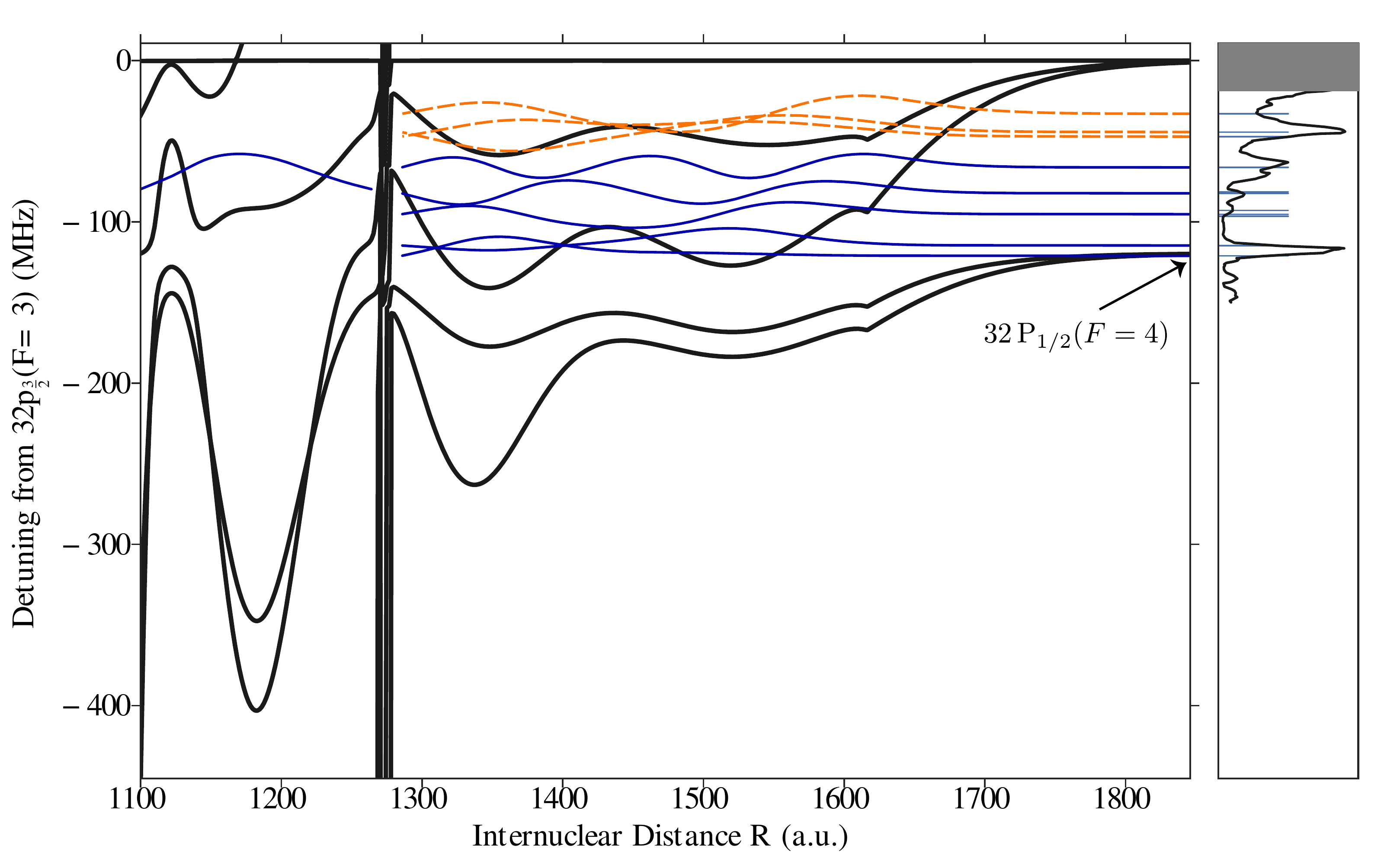}
\end{minipage}

\begin{minipage}{.5\textwidth}
  \centering
  \includegraphics[width=\linewidth]{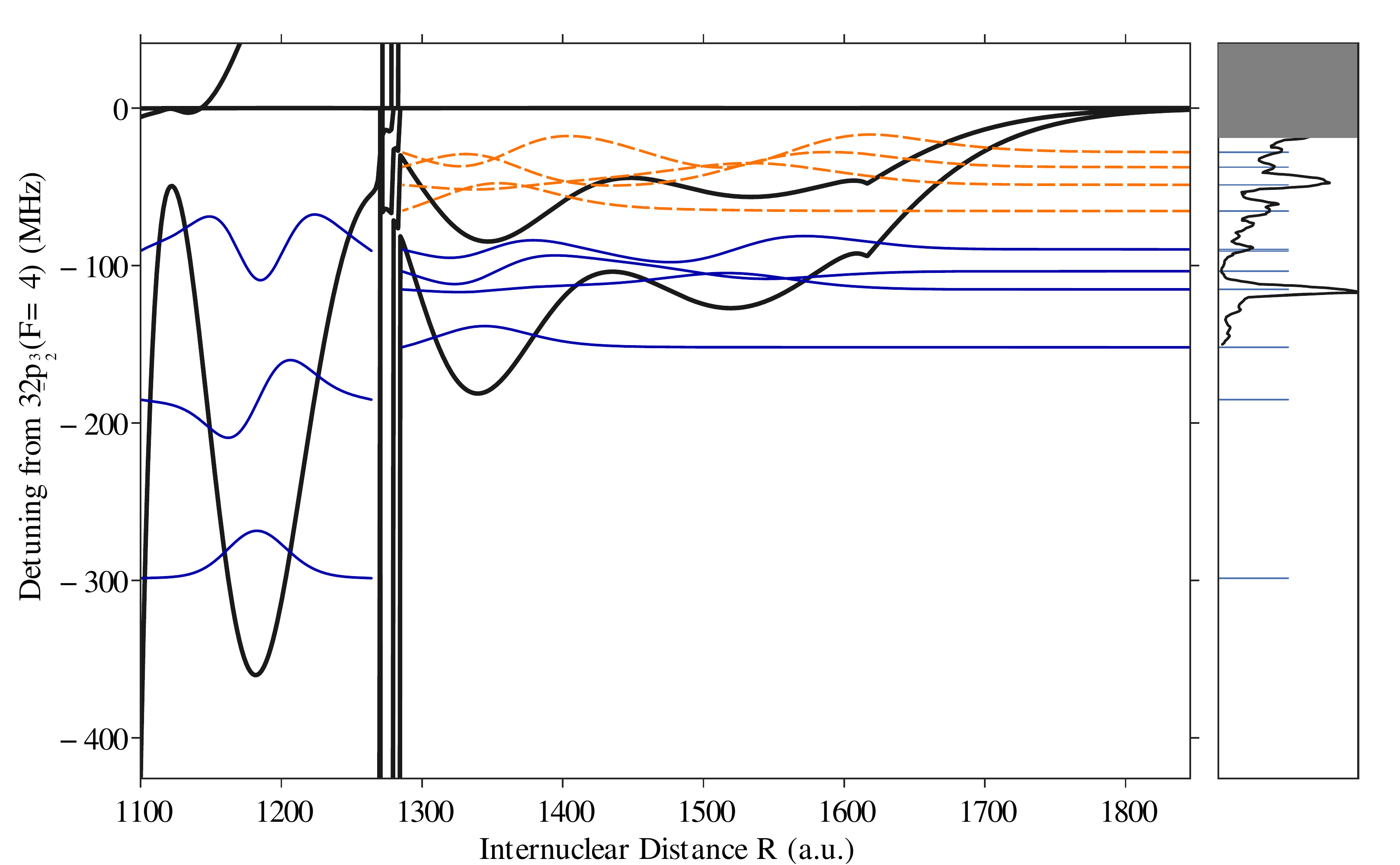}
\end{minipage}%
\begin{minipage}{.5\textwidth}
  \centering
  \includegraphics[width=\linewidth]{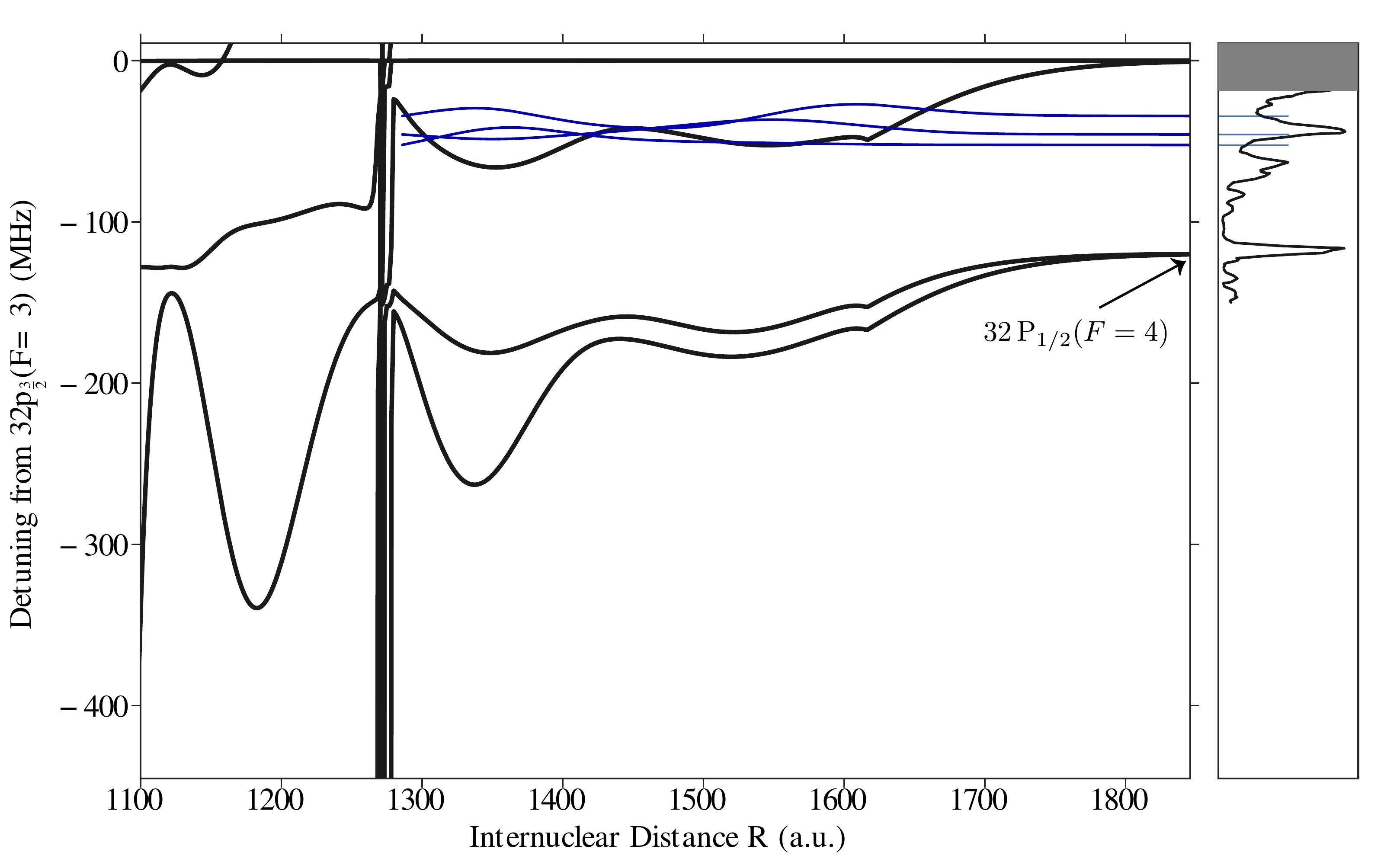}
\end{minipage}
\begin{minipage}{.5\textwidth}
  \centering
  \includegraphics[width=\linewidth]{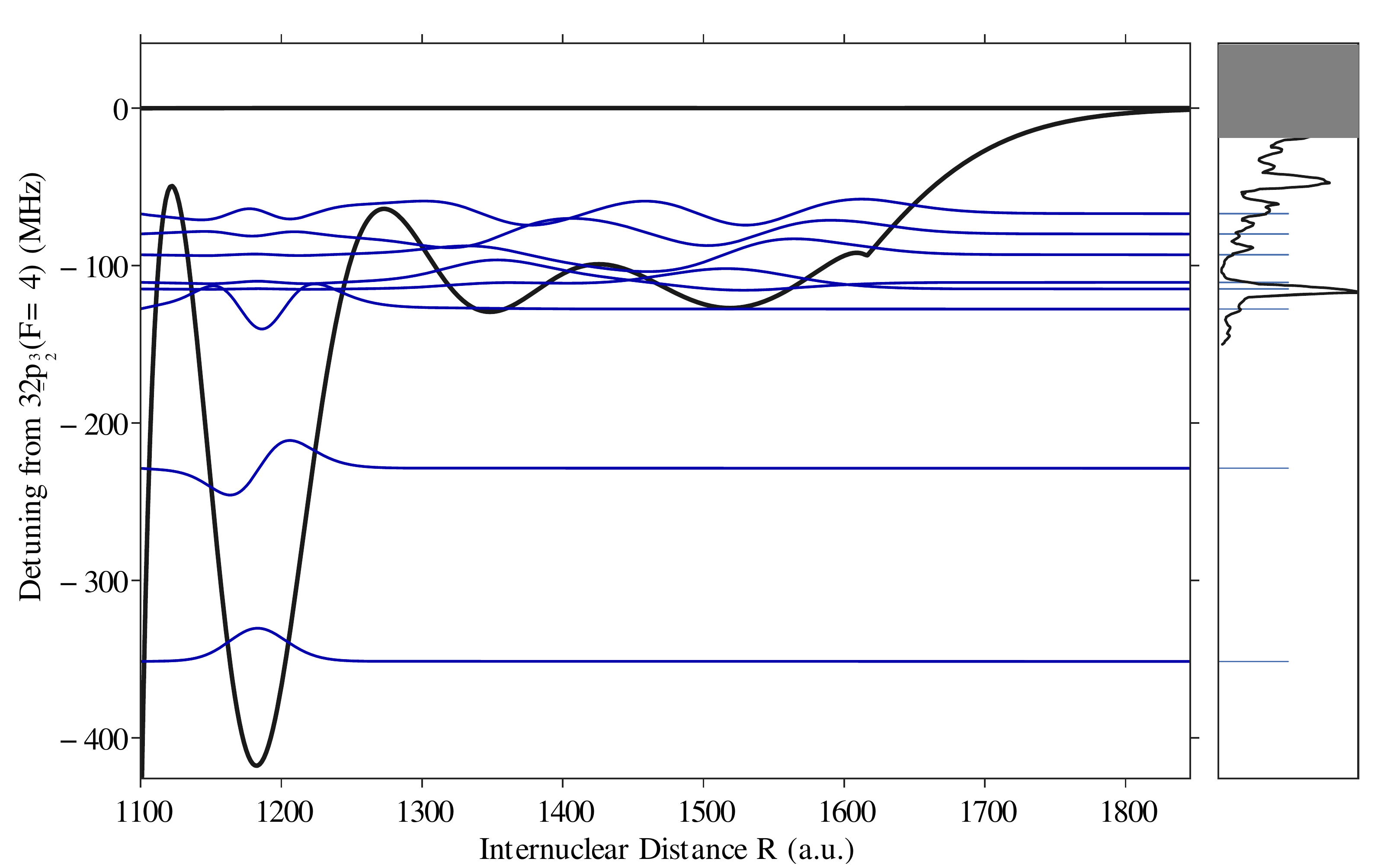}
\end{minipage}%

\caption{The BO potential energy curves correlating to the Cs$(6s)$ - Cs($32p_{3/2})$ F=3 (right) and F=4 (left and bottom) asymptotes, for $m_K=\frac{3}{2}$, $m_K=\frac{5}{2}$, $m_K=\frac{7}{2}$, $m_K=\frac{9}{2}$ (top to bottom). For $m_K=\frac{9}{2}$, the F=3 potential energy curves are not binding, being primarily of $\Pi$ molecular symmetry.}
\label{fig:mknon12}
\end{figure*}

The BO potential energy curves correlating to the Cs$(6s$) - Cs($32p_{3/2}$) F=3 and 4 asymptotes for  $m_K=\frac{3}{2}$,
 $m_K=\frac{5}{2}$,  $m_K=\frac{7}{2}$, and  $m_K=\frac{9}{2}$, are shown in Fig.~\ref{fig:mknon12}. The corresponding molecular bound states lead to additional spectral features that can be identified by comparing the vibrational energies to the observed spectral features. For $m_K=\frac{9}{2}$, the $F=4$ potential energy curves are not binding, and for the $F=3$ potential energy curves, there is no contribution from the $^3P_0$ scattering resonance, and hence the absence of any sharp avoided crossings.

\paragraph*{\textbf{  Spin weighting.}---} To accurately calculate the transition rates in photoassociation of trapped gas atoms into Rydberg molecules, we must take into account the electronic transition dipoles, as well as both the nuclear Franck-Condon and spin-overlap integrals.  
We assume that the atom pairs, which will bind into a Rydberg molecule,  are initially in states $\ket{6s(F_{r_0 }=\bar{F}),6s(F_{g_0}=\bar{F})}$,
i.e. the ground state atoms are in the same hyperfine state $\bar{F}$ when optically pumped \cite{sasmannshausen_experimental_2015}, and $F_{r_0}$($F_{g_0}$) refers to the hyperfine state of the ground state atom which will be Rydberg excited (will remain in the ground state). 
This initial state will be mixed uniformly and incoherently about $m_{F_r},m_{F_g} \in \{-\bar{F},\cdots,\bar{F}\} $.
For a given initial $m_{F_r},m_{F_g}$, the oscillator strength between two electronic states at  internuclear distance $R$ is proportional to
\begin{equation}
\bra{6s(F_{r_0},m_{F_{r_0}}),6s(F_{g_0},m_{F_{g_0}})} \hat{d} \ket{\psi}
\label{eq:fulloverlap}
\end{equation}
where $\ket{\psi}=\sum_i a_i \ket{\phi_i}$ are the calculated electronic eigenstates, and $\hat{d}$ is the electric dipole operator.

The overlap integral in Eq.~\ref{eq:fulloverlap} then becomes
\begin{multline}
O(R; m_{F_{r_0}}, m_{F_{g_0}})=\\
\bra{6s\cdots} \hat{d} \ket{\psi(R)} \braket{F_{r_0} m_{F_{r_0}} | \psi(R)} \braket{F_{g_0} m_{F_{g_0}} | \psi(R)} .
\end{multline}
We neglect in $\bra{6s\cdots} \hat{d} \ket{\psi}$ the contributions to $\ket{\psi}$ other than the 32p state.

For experimentally realized temperatures, the wave function of the ground state atom pair is constant on the scale of the Rydberg molecule wave function. Therefore the vibrational Franck-Condon factors take the form,  
\begin{equation}
FC_\nu(m_{F_{r_0}}, m_{F_{g_0}})\propto \int dR R^2 \psi_{\nu}(R) O(R;m_{F_{r_0}}, m_{F_{g_0}})
\end{equation}
for a given vibrational state $\psi_{\nu}(R)$.  

The overall Rydberg molecule formation rate is an incoherent sum over these Franck-Condon factors
\begin{equation}
\Gamma_\nu \sim \sum_{m_{F_{r_0}}} \sum_{m_{F_{g_0}}}| FC_\nu(m_{F_{r_0}}, m_{F_{g_0}})|^2G_L,
\end{equation}
where $G_L$ accounts for the laser profile.
The calculated absorption line profiles are compared with the observed spectra in Fig.~\ref{fig:fullspectra}, where a Gaussian laser line profile $G_L$ of width 5 MHz was used \cite{sasmannshausen_experimental_2015}. The agreement with the measured spectra (dashed lines)  is excellent. In this comparison the zero-energy triplet s-wave scattering length was adjusted by 5\%, i.e. $a_s^T(0) = - 20.71 \ a_0$.

\paragraph*{\textbf{ Electric dipole moments.}---} Our approach also allows for the prediction of electric dipole moments of Rydberg molecules. For the electronic wave functions  the transition and permanent electric dipole moments are
\begin{equation}
\bra{\psi} \hat{d} \ket{\psi} = \sum_i \sum_j \bra{\phi_i} \hat{d} \ket{\phi_j} a_i a_j .
\end{equation}
Because of the mixing of the opposite parity $(n-1)d$ and $(n-3)l_r\geq 3$ states with $np$ states in Cs, the Rydberg molecule obtains a permanent electric dipole moment (PEDM) \cite{li_homonuclear_2011,booth_production_2015}.
We note however that the dominant electronic transition is between the 32p and 31d atomic states whose dipole moment is $\bra{32p}\hat{d}\ket{31d}= 1583$ D;  
we neglect therefore all other contributions to the dipole moments. 
Thus, the spin-dependent dipole moments are 
\begin{equation}
d^S(R)= \sum_i \sum_j \bra{\phi_i} \hat{d} \ket{\phi_j} \delta_{m_{s_{r_i}},m_{s_{r_j}}} \delta_{m_{s_{g_i}},m_{s_{g_j}}} \delta_{m_{i_{g_i}},m_{i_{g_j}}}
\end{equation}
and the vibrationally averaged PEDM are
\begin{equation}
d^S_\nu = \int dR \psi_{\nu}^{\ast} (R) \psi_{\nu}(R) d^S(R).
\end{equation}
The PEDMs are calculated for the predominantly triplet ($d^T_\nu)$ and mixed ($d^{S+T}_\nu$) $\nu=0$ vibrational levels for each $m_K$ value. For the $m_K=\frac{1}{2}$  potential dissociating to the J=3/2, F=3(F=4) threshold the dipole moments are  $d^T_0$ = 9.8(8.5) D and $d^{S+T}_0$ = 3.7(3.7) D. For $m_K=\frac{3}{2}$, $d^T_0 $=  9.1(8.6) D and $d^{S+T}_0$ = 1.7(3.2) D, while for  $m_K=\frac{5}{2}$,  $d^T_0 $ = 7.8(8.2) D and $d^{S+T}_0$= 2.3(3.6) D, respectively.

\section{Outlook}

The level of spectroscopic precision of current experiments on Rydberg molecules combined with spin resolution allow for the deterministic evaluation of such fundamental properties  as the scattering length and  energies of scattering resonances of electron-neutral atom collisions.  The spin-dependence of the Rydberg molecular potentials opens the possibility to control the spin state of the molecules and to realize paradigm spin models in an entirely new regime. Our method can be readily extended to study such phenomena as spin and angular momentum alignment in high angular momentum Rydberg molecules such as Cs(nd).

\section{Acknowledgements}
The computations in this paper were run on the Odyssey cluster supported by the FAS Division of Science, Research Computing Group at Harvard University.
\noindent S.M. is supported by an NSF Graduate Research Fellow grant, the Institute for Theoretical Atomic, Molecular, and Optical Physics at Harvard University, and the Smithsonian Astrophysical Observatory.
\noindent R.S. is supported by the NSF through a grant for the Institute for Theoretical Atomic, Molecular, and Optical Physics at Harvard University and the Smitsonian Astrophysical Observatory.  
\noindent S.T.R. acknowledges support from NSF Grant No. PHY-1516421 and from a Cottrell College Science Award through the Research Corporation for Scientific Advancement.
\noindent J.S. acknowledges support from NSF Grant No. PHY-HY-1205392.

%\bibliography{xfine}


\end{document}